\documentclass[11pt,leqno,twoside]{article}

\RequirePackage[OT1]{fontenc}
\RequirePackage{amsthm,amsmath,fancyhdr}
\RequirePackage[authoryear]{natbib}
\usepackage{mathrsfs,amssymb,amsfonts,amsthm,amsmath,natbib,amsbsy,dsfont}
\usepackage[ruled,vlined]{algorithm2e}
\usepackage{graphicx,float,subfigure,booktabs}
\usepackage[usenames,dvipsnames]{xcolor}
\usepackage[bookmarks,bookmarksnumbered,colorlinks=true,pdfstartview=FitV, linkcolor=Red,citecolor=blue!90!green,urlcolor=blue!90!green]{hyperref}

\allowdisplaybreaks

\bibpunct{\textcolor{blue}{(}}{\textcolor{blue}{)}}{;}{a}{\textcolor{blue}{,}}{\textcolor{blue}{;}}

\long\def\symbolfootnote[#1]#2{\begingroup\def\thefootnote{\hspace*{-1mm}\fnsymbol{footnote}}\footnote[#1]{#2}\endgroup}

\def\ind{\mathds{1}}

\def\maxy{B}
\def\d{\mathrm{d}}
\def\X{\mathbb{X}}

\def\N{\mathbb{N}}
\def \R{\mathbb{R}}

  \DeclareMathOperator*{\argmin}{arg\,min}

\numberwithin{equation}{section}
\theoremstyle{plain}

\newtheorem{theorem}{Theorem}[section]

\newtheorem{definition1}[theorem]{Definition} 
\newtheorem{remark1}[theorem]{Remark}

\title{\vspace{-15mm}\bf  
Bayesian nonparametric forecasting of monotonic functional time series}
\author{
\textsc{Antonio Canale}\\
\emph{University of Torino and Collegio Carlo Alberto}\\[1mm]
\textsc{Matteo Ruggiero}\\
\emph{University of Torino and Collegio Carlo Alberto}\\[1mm]
}
\date{\today}

\fancypagestyle{plain}{
\fancyhead{}
}

\begin{document}

\maketitle
\thispagestyle{empty}

\begin{abstract}
We propose a Bayesian nonparametric approach to modelling and predicting a class of functional time series with application to energy markets, based on fully observed, noise-free functional data. Traders in such contexts conceive profitable strategies if they can anticipate the impact of their bidding actions on the aggregate demand and supply curves, which in turn need to be predicted reliably. Here we propose a simple Bayesian nonparametric method for predicting such curves, which take the form of monotonic bounded step functions. We borrow ideas from population genetics by defining a class of interacting particle systems to model the functional trajectory, and develop an implementation strategy which uses ideas from Markov chain Monte Carlo and approximate Bayesian computation techniques and allows to circumvent the intractability of the likelihood. Our approach shows great adaptation to the degree of smoothness of the curves and the volatility of the functional series, proves to be robust to an increase of the forecast horizon and yields an uncertainty quantification for the functional forecasts.
We illustrate the model and discuss its performance with simulated datasets  and on real data relative to the Italian natural gas market. \medskip

\medskip

\end{abstract}


\section{Introduction and motivation}

Many radical changes recently occurred in several European energy markets have determined new logistic, economic and statistical challenges. In the Italian natural gas market, for example, new regulations have introduced daily auctions to balance the common pipeline network on a virtual balancing platform. This operative mechanism produces datasets of functional time series, where single data points consist in monotonic step functions that represent daily demand and supply curves. Such functions result from appropriately sorting all the operators' information and have random number, size and location of jumps. Market traders are then interested in predicting the effect of their own bidding strategies on the exchange price forecast, which is evaluated by manipulating the curves and cannot be obtained similarly by exploiting univariate price prediction. In Section \ref{sec:context} we provide a stylised illustration of this point. 
Being able to predict and to quantify the uncertainty concerning the next day's curves is thus of dramatic interest both for common market operating strategies as well as for speculative reasons. 
In this framework, new economic opportunities call for suitable inferential methods for the analysis of functional time series, with particular reference to $h$-step-ahead functional forecasting.

A specificity of the present framework is the type of data we consider,  which form a noise-free functional time series. This gives rise to intrinsically infinite-dimensional observations. 
Here we propose a Bayesian nonparametric approach for modelling and forecasting these monotonic bounded step functions. 
Unless one is willing to impose unrealistic constraints, the data at hand cannot be efficiently described by a parametric model, since the number, size, and location of the jumps are to be considered as random. A nonparametric approach is therefore needed, with an underlying diffuse distribution modelling the jump locations and discrete random probability measures modelling the jump sizes. Borrowing ideas from interacting particle systems, we model the transformed offer and inverse-demand functions by resorting to a latent Markov chain of interacting particles and then considering the time series of cumulative distribution functions (cdf) of the particles. The approach is able to provide both point estimates and uncertainty quantifications for general $h$-step-ahed functional predictions. Motivated by the same application \cite{CV16} recently proposed a frequentist method based on functional autoregression techniques, following the classical approach to functional data analysis of \citet{RS05}. However, the latter approach is limited to one-step-ahead forecasting and does not provide any measure of uncertainty.

Bayesian nonparametric modelling in dynamic frameworks has received considerable attention recently, inspired by the ideas first proposed by \cite{ME99,ME00} on the so-called dependent processes. These can be generally formulated as a collection of discrete random probability measures $P_{z}=\sum_{i\ge1}p_{i,z}\delta_{X_{i,z}^{*}}$
where the weights $p_{i,z}$ and/or the atoms locations $X_{i,z}^{*}$ are indexed by a covariate $z$, which can account for time dependence.  We refer the reader to \cite{HHMW} for reviews and references. Among dependent processes indexed by time, \cite{D06} models the dependent process as an autoregression with Dirichlet distributed innovations; \cite{GS10} reduce the innovation to a single atom sampled from the centering measure; 
\cite{Cea06} model the noise in a dynamic linear model with a Dirichlet process mixture; \cite{RT08} who induce the dependence in time only via the atoms, by making them into an heteroskedastic random walk; \cite{MRW11} construct a dependent model with geometric weights; \cite{MR16} define a diffusive Dirichlet mixture with Wright--Fisher weights. Other contributions that exploit Polya urns for constructing dependent models are \cite{CDD07} and \cite{PR13}. See also \cite{Fea13} for a unifying representation of some of the  cited dependent random processes.

As mentioned above, here the nature of the data differs from typically approached with Bayesian nonparametric mixture modelling, where one is given a vector of observed vector-valued data points $y$, assumed to be generated by a density supported on a finite-dimensional space. A Bayesian nonparametric approach to such problem can for instance assign a prior distribution to the latent distribution of the parameters, and then condition on the data to yield, typically with the aid of Markov chain Monte Carlo (MCMC) strategies, inferences on the generating density. In our framework, however, the daily single data point consists instead in a pair of demand and supply curves which are monotone bounded step function with random number, size and location of the jumps. Therefore here the data 
 form a noise-free functional time series, taking values in an infinite-dimensional space and with intractable likelihood. This in turn prevents from adopting a usual conditioning argument on the available data, not allowing to devise common MCMC inferential strategies aiming at the latent process that generates the curves, but requiring instead a  likelihood-free approach to posterior inference.

We address this goal by borrowing ideas from population genetics and constructing a class of interacting particle systems which, together with a suitable computational strategy, lend themselves to our inferential purposes. More specifically, we induce a functional time series by means of an underlying latent particle Markov chain. The dynamics of the particles are induced by means of P\'olya urn type updates applied to a random fraction of the population of particles. These features allow the model to efficiently capture the latent distribution of the jump locations, to adapt to the (functional) volatility of the time series and to learn the degree of smoothness of the sample path. 
The features of the resulting dependent Dirichlet process model, which shares some points with the approaches in \cite{CDD07} and \cite{FRW09}, are tailored to the application at hand and lend nicely to an alternative inferential strategy. We devise a suitable algorithm for posterior computation which exploits ideas from MCMC and approximate Bayesian computational (ABC) methods. This circumvents the intractability of the implied likelihood and allows to estimate the latent tracts of the data generating mechanism.

Adopting a Bayesian nonparametric approach jointly with using interacting particle systems is advantageous in this framework. Bayesian nonparametric temporal modelling has shown to provide great flexibility if compared with classical parametric approaches, since it enables the model to capture elaborate data dynamics whilst preserving a relative ease in the necessary computation. Using interacting particle systems in this framework allows to tune and learn the above mentioned structural features of the time series, trading off between flexibility and precision and keeping the modelling machinery light and simple. Furthermore, a Bayesian approach has the advantageous byproduct of automatically providing a measure of the predictive uncertainty, which is not available with other approaches to the same problem. Quantifying the forecast uncertainty is a crucial issue, for this application and more generally in functional data analysis, still debated in the realm of frequentist approaches \citep{RS05}.

The rest of the paper is organised as follows. In Section \ref{sec: model}, we introduce a 
 dependent process for monotonic step functional times series modelling. Section \ref{sec: abc} outlines the strategy for posterior computation.
Section \ref{sec: sim} presents a simulation study for testing the algorithm's performance in predicting future curves and in learning the underlying generating process. In Section \ref{sec: appl} we implement the model to the energy market dataset and discuss our findings. Section \ref{sec: extension} concludes with some brief remarks.


\section{A dependent model for functional forecasting}\label{sec: model}

We assume the data take the form of a time series of step functions
\begin{equation}\label{data}\nonumber
F=\{F_{t}(\cdot)\}_{t=0,\ldots,T},\quad \quad F_{t}(\cdot):\R \rightarrow [0,\maxy],\quad \quad T, B>0,
\end{equation} 
where each $F_{t}(\cdot)$ is right-continuous, non decreasing, and takes on finitely-many values.  Here, for simplicity, we assume $T,B<\infty$, which is justified by the energy market application, and without loss of generality we can therefore take $B=1$. Hence the data can be thought of as a cdf-valued time series, where each time instance has random number, size and locations of the jumps. Note that the latent infinite-dimensional parameter in usual nonparametric mixture models is instead here the observable.

We model the data by means of a latent particle system $\{X^{(n)}(t)\}_{t=0,\ldots,T}$, where $X^{(n)}(t)=(X_{1}(t),\ldots,X_{n}(t))$ is a vector of interacting $\R$-valued, discrete-time Markov chains. Given the trajectory of  $\{X^{(n)}(t)\}$ we induce a cdf-valued Markov chain by defining
\begin{equation}\label{curve t-indexed}
F_{t}(x)=
\frac{1}{n}\sum_{i=1}^{n}\ind(X_{i}(t)\le x).
\end{equation} 
The particle dynamics are defined as follows. At each discrete time $t\in\N$, given $X^{(n)}(t-1)$, the next state $X^{(n)}(t)$ is obtained as follows:
\begin{itemize}
\item sample $M\sim \text{Binom}(n,p)$;
\item given $M=m$, choose $m$ indices $i_{1},\ldots,i_{m}$ in $\{1,\ldots,n\}$ without replacement;
\item replace the coordinates in $X^{(n)}$ indexed by $i_{1},\ldots,i_{m}$ with an $m$-sized sample from a Blackwell-MacQueen Polya urn with total mass parameter $\theta$ and base measure $P_{0}(\cdot)$, conditionally on the $(n-m)$ remaining particles. More specifically, sample
\begin{equation}\label{resampling}
\begin{split}
X_{i_{1}}\sim&\,\frac{\theta}{\theta+n-m}P_{0}(\cdot)+\frac{1}{\theta+n-m}\sum_{j\ne i_{h}:h=1,\ldots,m}\delta_{X_{j}}(\cdot)\\
X_{i_{2}}\sim&\,\frac{\theta}{\theta+n-m+1}P_{0}(\cdot)+\frac{1}{\theta+n-m+1}\sum_{j\ne i_{h}:h=2,\ldots,m}\delta_{X_{j}}(\cdot)\\
\vdots&\\
X_{i_{m}}\sim&\,\frac{\theta}{\theta+n-1}P_{0}(\cdot)+\frac{1}{\theta+n-1}\sum_{j\ne i_{m}}\delta_{X_{j}}(\cdot),
\end{split}
\end{equation} 
\item add to the vector the $(n-m)$ remaining particles.
\end{itemize}
A special case of the above construction has been used in \cite{RW09a, RW09b} as a building block for studying the asymptotic properties of certain processes of interest in Mathematical Biology. In particular it can be seen as a modification of a Moran type particle system, a Markov chain model which has long been known in Population Genetics. This describes a population with overlapping generations (here given by the discrete time steps), with each generation formed by the previous where one individual has been substituted according to a fixed distribution. An instance of Moran model can thus be obtained by  fixing $M=1$. See, e.g., \cite{E09}. 
By exploiting the exchangeability of a P\'olya urn sequence, one can show that the Moran model is stationary and reversible with respect to the joint law of a P\'olya urn sample (see, e.g., \citealp{RW09b}). Our construction then is equivalent to an accelerated Moran process, where the population is observed at random intervals of generations, and can  be shown to share the same properties. In Section \ref{sec: extension} some possible extensions of the present model, including a non-Markovian version, will be briefly outlined. 

The roles of the model parameters on the induced dynamics can be described as follows. 
The concentration parameter $\theta$ and the base distribution $P_{0}$ control the marginal properties of the underlying collection of random probability measures, in particular $P_{0}$ the shape and $\theta$ degree of smoothness of each curve, the latter increasing with $\theta$. Jointly, $\theta$, $n$, and $p$ control the dynamic properties by regulating how close successive curves are on average  and the speed of innovation of the jump locations within each curve. The parameter $\theta$ plays a key role in the resampling mechanism \eqref{resampling}, by determining the probability of introducing new jump locations as opposed to reweighing current locations, and by determining, jointly with $p$, the amount of shared locations between successive curves and thus the closeness between curves. 
On the other hand, $n$ and $p$ trade-off between rougher and smoother dynamics. When $p=0$, no particles are updated, and the model reduces to a static $n$-sized sample from a Dirichlet process. When $p=1$, the entire population of particles is updated at every transition, and the qualitative  effects of this renewal depend on the other parameters. In the latter case, the particle dynamics are asymptotically equivalent, for large $n$, to a Wright--Fisher process with infinitely-many types, as that used in \cite{MR16}. Intuitively, this is due to the fact that the transition determined by $n$ generations of a Moran model is close in distribution with that determined by one generation of a Wright--Fisher process, ultimately due to the closeness of repeated Polya urn updating to Multinomial sampling with frequencies biased by a mutation mechanism. See, e.g., \cite{E09} for more details on these connections.

The model is completed by assigning suitable prior distributions to $\theta$, $p$ and $\vartheta$, the latter being an $\R^{d}$-valued vector of parameters that characterises the distribution $P_{0}=P_{0}(\cdot\mid \vartheta)$. For the time being, denote such prior by $\pi$, i.e.
\begin{equation}\label{hyperpriors}
\eta = (\theta, p, \vartheta) \sim \pi(\eta)=\pi_{1}(\theta\mid \sigma_{1})\times \pi_{2}(p\mid \sigma_{2})\times \pi_{3}(\vartheta\mid \sigma_{3})
\end{equation} 
where for simplicity we have assumed independence, and $\sigma_{i}$ denotes a vector of hyperparameters.

\section{Posterior computation and predictive inference}\label{sec: abc}

\subsection{General strategy}

Given the nature of the available data is not that typically available in Bayesian nonparametric mixture modelling, the framework does not allow for common MCMC strategies. Lacking a tractable likelihood for the functional time series prevents from using usual conditional arguments and requires a different, likelihood-free approach. Here we approach posterior inference by combining ideas from Markov chain Monte Carlo and approximate Bayesian computation (ABC). ABC methods are computational techniques that do not rely on the availability of a likelihood and have been successfully used in the past decade among the most satisfactory approaches to intractable likelihood problems. The basic ABC idea is to generate a candidate parameter from the prior, draw an observation from the sampling model conditionally on the candidate parameter and  measure an appropriate distance between synthetic and real data, typically done by means of suitable summary statistics. The candidate parameter is retained as a sample from the posterior distribution if the distance is less than a specified acceptance threshold. See \citet{Mea12} for a recent review. 

Here we adapt an idea of \cite{Marj03} to the present setting by adding an MCMC step in the ABC strategy which improves the efficiency of the algorithm. Inspired by Algorithm~3 in \cite{Mea12}, we use a standard ABC routine for generating the first instance of accepted parameters and then switch to an ABC criterion based on a Metropolis--Hastings proposal distribution. More specifically, we generate candidate parameters $\eta = (\theta,p,\vartheta)$ from \eqref{hyperpriors} and then generate a  sample path $F^{*}=\{F^{*}_{t}(x\mid \eta)\}_{t=0,\ldots,T}$ from the model, conditional on the candidate vector $\eta$ (henceforth we suppress the dependence of $F^{*}_{t}$ and similar quantities on $\eta$ for notational simplicity). The candidate $\eta$ is accepted if the proposal sample path satisfies a distance condition
\begin{equation}\label{distance general}
d(\rho(F^{*}),\rho(F))\le\varepsilon,
\end{equation} 
where $F$ is the available data as in \eqref{data}, $\rho$ is an appropriate summary of the functional sample path, $d$ is a suitable distance and $\varepsilon$ is a chosen threshold. Given the parameters have different roles and implications on the model properties, as discussed in Section~\ref{sec: model}, 
here we choose to specialise \eqref{distance general} into $3$ different conditions, one for each of $\theta$, $p$  and the baseline Dirichlet parameters $\vartheta \in \R^{d}$. This is motivated with the observation that $\vartheta$ mainly affects the marginal properties of the model, $p$ the dynamic properties, and $\theta$ both. Therefore we set
\begin{equation}\label{distances}
d_{j}(\rho_{j}(F^{*}),\rho_{j}(F))\le\varepsilon_{j},\quad j=1,2,3,
\end{equation} 
where $d_{j},\rho_{j},\varepsilon_{j}$ specify distances, summaries and thresholds for each of $\theta$, $p$, and $\vartheta$, to better capture the different features of the samples paths for which they are responsible. In this framework, the distances are related to the average group clustering of the latent particles, to the volatility of the functional series, and to the curves shape, respectively. These are discussed in detail in the following Section \ref{sec: distances}.

Upon acceptance of the first candidate, denoted $\eta^{(1)}$, we propose new candidate values with the random-walk Metropolis--Hastings kernel
\begin{equation}\label{proposal kernel}
\eta^{*}\sim q(\eta^{*}\mid \eta^{(i)}), 
\end{equation} 
where $\eta^{(i)}$ is the last accepted value, and generate a new sample path $F^{*}$ conditional on $\eta^{*}$. If the Metropolis--Hastings condition
\begin{equation}\label{metropolis condition}
u\le \frac{\pi(\eta^{*})q(\eta^{(i)}\mid \eta^{*})}
{\pi(\eta^{(i)})q(\eta^{*}\mid \eta^{(i)})},\quad \quad u\sim \text{Unif}(0,1), 
\end{equation} 
is satisfied, together with the conditions \eqref{distances}, then $\eta^{(i+1)}:=\eta^{*}$ is accepted, otherwise $\eta^{(i+1)}:=\eta^{(i)}$. 
The resulting sample $(\eta^{(i)},i=1,\ldots,I)$ is an $I$-sized approximate draw from the posterior distribution of the parameters. Algorithm~\ref{algorithm} summarises the pseudo-code of the above strategy.

\IncMargin{1.6em}
\begin{algorithm}[t]
\vspace{2mm}
\hspace{-10mm} \KwData{Step functions $F=\{F_{t}(x)\}_{t=0,\ldots,T}$}
\hspace{-10mm} Set prior hyperparameters $\sigma_{i}$, $i=1,2,3$\\
\hspace{-10mm} Generate first accepted vector by\\
\Repeat{$d_{j}(\rho_{j}(F^{*}),\rho_{j}(F))\le\varepsilon_{j}$ \emph{\textbf{for all}} $j=1,2,3$}{
sample $\eta^{*}=(\theta,p,\vartheta)$ from prior $\pi$\\
generate $F^{*}=\{F^{*}_{t}(x)\}_{t=0,\ldots,T}$ given $\eta^{*}$}
Set $\eta^{(1)}=\eta^{*}$\\
\hspace{-10mm} Generate an $I$-sized MCMC-ABC sample by

 \For{$i=2,\ldots,I$}{
 sample $\eta^{*}\sim q(\cdot \mid \eta^{(i-1)})$\\
 generate $F^{*}=\{F^{*}_{t}(x)\}_{t=0,\ldots,T}$ given $\eta^{*}$\\
  sample $u\sim \text{Unif}(0,1)$
  
   \eIf{$u\le \displaystyle \frac{\pi(\eta^{*})q(\eta^{(i-1)}\mid \eta^{*})}
{\pi(\eta^{(i-1)})q(\eta^{*}\mid \eta^{(i-1)})}$ \emph{\textbf{and}} 
$d_{j}(\rho_{j}(F^{*}),\rho_{j}(F))\le\varepsilon_{j}$ \emph{\textbf{for all}} $j=1,2,3$}{
  	$\eta^{(i)}:=\eta^{*}$
   }{
  	$\eta^{(i)}:=\eta^{(i-1)}$
  }
 }
\hspace{-10mm} \KwResult{Approximate sample $\{\eta^{(i)},i=1,\ldots,I\}$ from posterior}
\vspace{2mm}
 \caption{MCMC-ABC algorithm for curves evolution}\label{algorithm}
\end{algorithm}

Once an MCMC-ABC posterior sample $(\eta^{(i)},i=1,\ldots,I)$ is available, this is used to generate the predictive estimate according to the model given the last data point $F_{T}(x)$. 
The $h$-step-ahead curve forecast can then be obtained as a Monte Carlo point estimate of the posterior $h$-step-ahead predictive mean
\begin{equation*}
f_{h}(x)=\int F_{T+h}(x)\, p(F_{T+h}(x)\mid \eta,F_{T})\hat\pi(\eta\mid F_{1,\ldots,T})\d \eta.
\end{equation*} 
Here $F_{1,\ldots,T}$ is the short notation for $\{F_{t}(x)\}_{t=1,\ldots,T}$, $p(F_{T+h}\mid \eta,F_{T})$ is the $h$-step-ahead transition function of the functional process and $\hat\pi(\eta\mid F_{1,\ldots,T})$ is the approximate posterior distribution of the parameters given the data, for which an ABC sample is available. Note, however, that given a predictive sample $\{f^{(i)}(\cdot)\}_{i=1,\ldots,I}$, the mean prediction evaluated pointwise
\begin{equation*}
\bar f(x)=\frac{1}{I}\sum_{i=1}^{I}f^{(i)}(x)
\end{equation*} 
does not possess qualitative features similar to the origina data, resulting in an oversmoothed curve. 
Then we let the point estimate $\hat f$ be
\begin{equation}\label{point estimate}
\hat f(x)= f^{(i^{*})}(x),\quad \quad 
i^{*}=\argmin_{i=1,\ldots,I}L_{2}(f^{(i)},\bar f),
\end{equation} 
that is $\hat f(x)$ is the Monte Carlo sample $f^{(i)}$ with minimum  $L_{2}$ distance from the pointwise mean predictive estimate $\bar f$. 
A similar approach to such issue is used in \citet{dahl} in the context Dirichlet process mixture clustering.

Posterior credible bands for the point estimate are also available based on the MC predictive sample. This uncertainty quantification on the point estimate is usually not readily available with frequentist strategies to the same problem, whereas it is a free byproduct of the Bayesian approach. 

\subsection{Distance criteria}\label{sec: distances}

We discuss here in more detail the distance criteria which seem fit for this problem. Denote by $F^{*}$ the trajectory simulated by the ABC sampler, and use the same superscript for the associated summaries. We specialise the distance criteria \eqref{distances} to the following:
\begin{equation}\label{soglie sim1}
\begin{aligned}
d_{1}(\rho_{1}(F^{*}),\rho_{1}(F))
=&\,|\bar K^{*}-\bar K|\le\varepsilon_{1},\\
d_{2}(\rho_{2}(F^{*}),\rho_{2}(F))
=&\,L_{2}(\bar F^{*},\bar F)\le\varepsilon_{2},\\
d_{3}(\rho_{3}(F^{*}),\rho_{3}(F))
=&\,\vert \bar L_{2}(F^{*})-\bar L_{2}(F)\vert\le\varepsilon_{3}.
\end{aligned}
\end{equation} 
The first is the absolute difference between the mean number of jumps $\bar K=T^{-1}\sum_{t=1}^{T}K_{t}$ in each times series, $K_{t}$ being the number of jumps at time $t$, which provides information on the average clustering of the latent particles. The second is the $L_{2}$ distance between the ergodic pointwise means of the two time series, where
\begin{equation*}
\bar F_{t}(x)=\frac{1}{T}\sum_{t=1}^{T}F_{t}(x). 
\end{equation*} 
This is a measures of closeness of the two time series and also provides information about the shape of the marginal distribution of the jump locations. The third is the absolute difference between the ergodic means of the $L_{2}$ distances between consecutive curves in each time series, where
\begin{equation*}
\bar L_{2}(F)=\frac{1}{T-1}\sum_{t=2}^{T} L_{2}(F_{t},F_{t-1}).
\end{equation*} 
This provides a measure of the volatility of each time series by comparing the average displacement between successive curves in each trajectory. All distances between curves are computed on a 500-length discrete grid for curves normalised to lie in $[0,1]$. 

The choice of thresholds in \eqref{soglie sim1} that yield satisfactory acceptance rates and exploration of the state space is clearly of great importance for the quality of the outcome, and constitute an open problem in ABC methodology. Indeed, few ABC strategies allow to invoke a general rule for choosing the acceptance thresholds, which must otherwise be calibrated. 

Here we propose a strategy to fixing the thresholds that exploits summary statistics of the data, which can in principle be used for different datasets. In particular we set
\begin{equation}\label{c-soglie sim1}
\begin{aligned}
\varepsilon_{1}=c_{1} \bar K,\quad 
\varepsilon_{2}=c_{2} L_{2}(F_{\max},F_{\min}),\quad 
\varepsilon_{3}=c_{3} \bar L_{2}(F),
\end{aligned}
\end{equation} 
where $c_{j}\in[0,1]$, $\bar K$ and $\bar L_{2}(F)$ are as above, and $L_{2}(F_{\max},F_{\min})$ is the $L_{2}$ area of the convex hull of the data, with $F_{\max}$ and $F_{\min}$ being the minimum majorant and the maximum minorant of the time series. 
By using summaries of the data closely related to those involved in the definition of the associated distances, one can interpret the thresholds as percentages of oscillation of the synthetic data, generated conditional on the candidate parameters, with respect to the real data. That is, $\varepsilon_{j}$ reflects a $c_{j}$\% oscillation of the sample generated by the candidate parameters with respect to the data, relative to distance $d_{j}$. 
Note that here we are letting the $c_{j}$'s be percentages, but ratios bigger than one can sometimes be useful depending on the data. In general we do not expect $c_{j}$ to have the same values for all $j$. In particular we expect $c_{2}$ to be sensibly different from $c_{1},c_{3}$, since $d_{1},d_{3}$ compare numbers whereas $d_{2}$ compares functions.


\section{Simulation study}\label{sec: sim}

We start by assuming the model is correctly specified, which allows a more thorough investigation of the performance of our estimating procedure, and later consider the case of misspecification. We first generate the data $F=\{F_{t}(x)\}_{t=0,\ldots,T}$ from the model of Section \ref{sec: model} with $\X=[0,1]$, $\theta=10$, $n=500$, $p=0.7$, $P_{0}=\text{Beta}(.25,.3)$ and time horizon $T=110$. These values have been taken to partially mimic some feature of the real dataset analyzed in Section~\ref{sec: appl}. 
\begin{figure}[t!]
\begin{center}
\includegraphics[width=.495\textwidth]{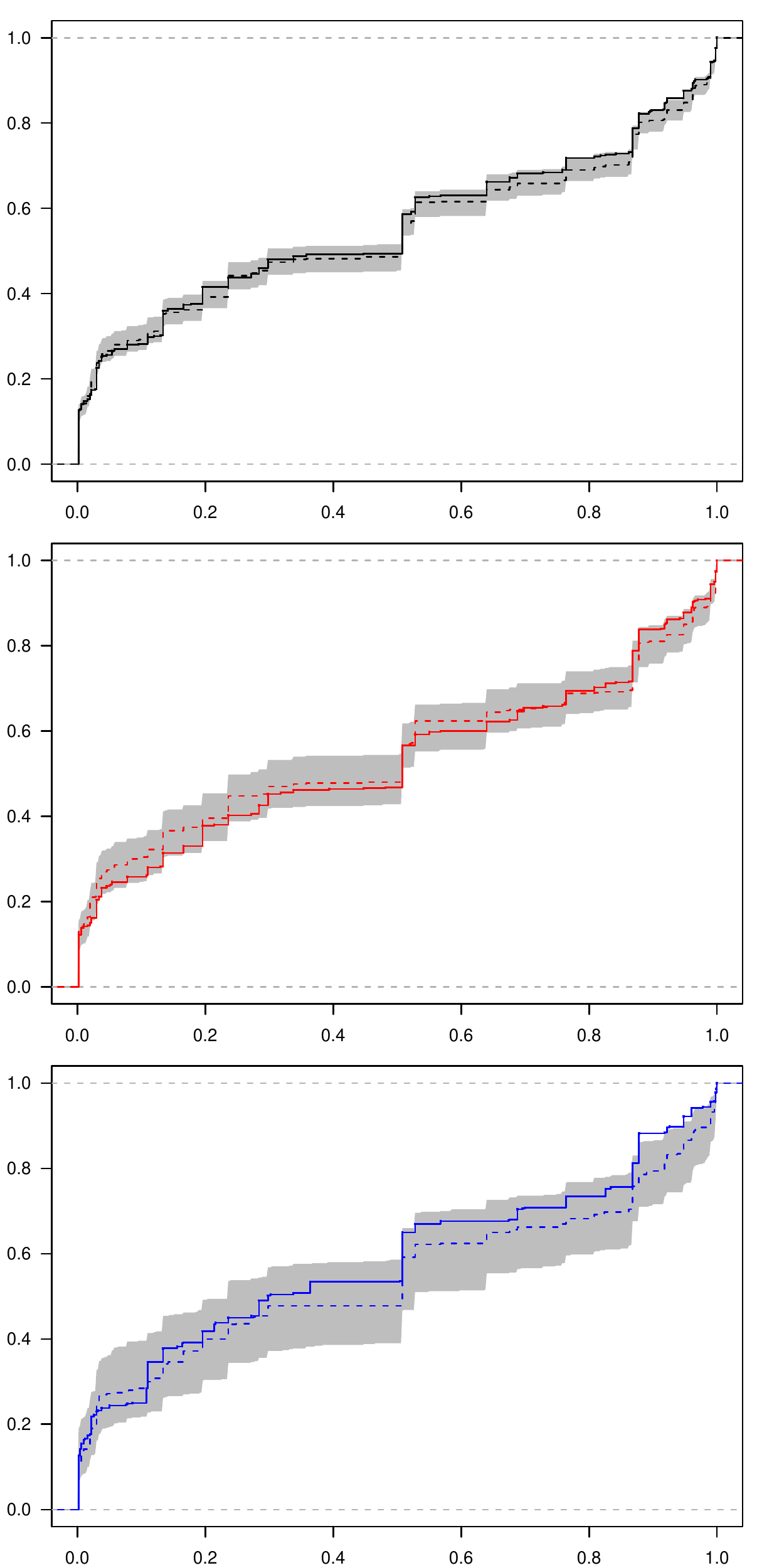}
\includegraphics[width=.495\textwidth]{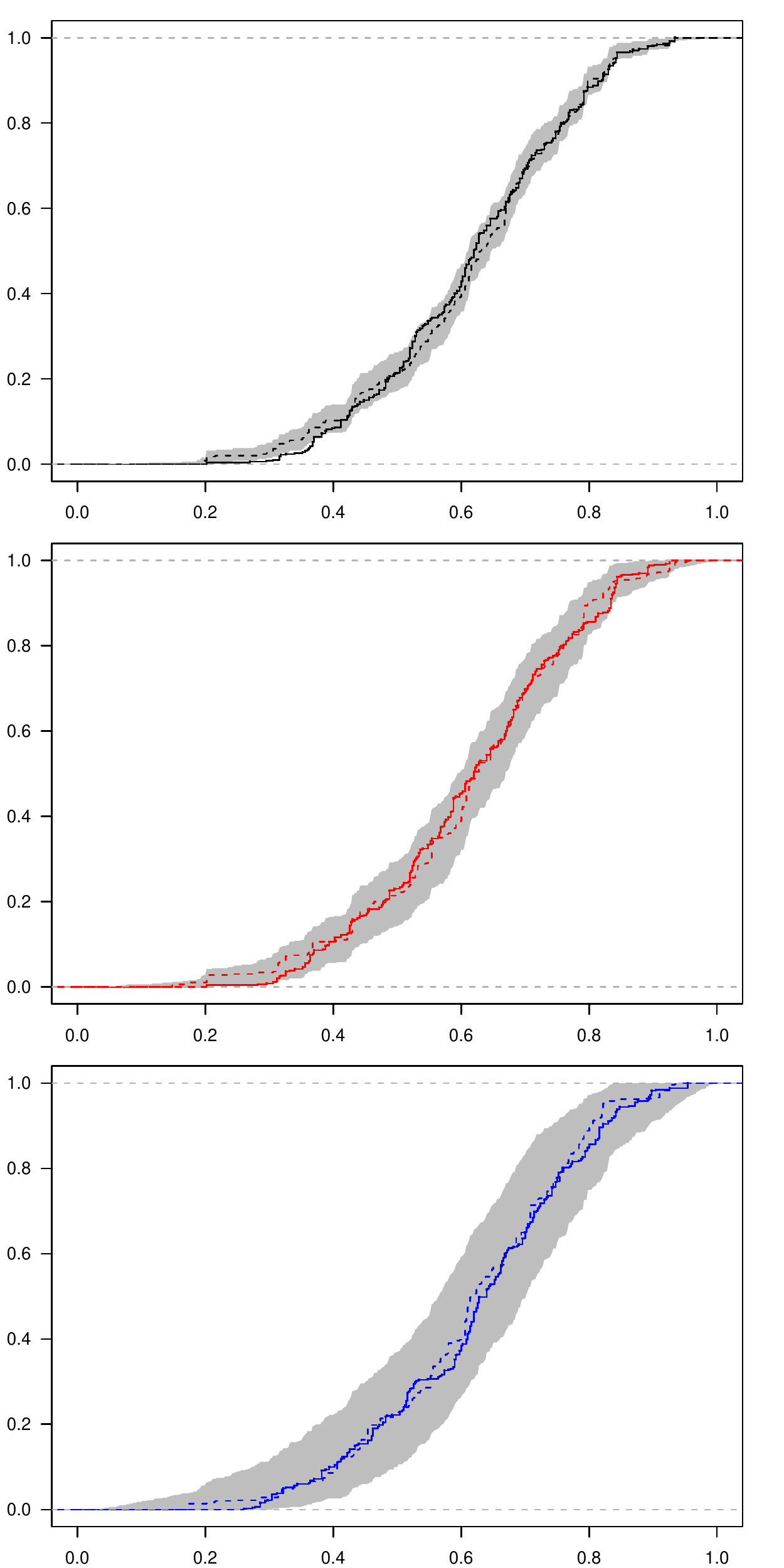}
\end{center}
\begin{quote}
\caption{\scriptsize 
$h$-step-ahead forecasts for the correctly specified model (left) and the misspecified model (right). Each picture shows the 99\% pointwise credible intervals (grey bands), the forecast point estimate (dashed line) and the true curve (solid line), for $h$ equal to 1 (top), 3 (middle) and 10 (bottom).
}\label{fig:sim1-band}
\end{quote}
\end{figure}
We complete the model specification by choosing 
$P_{0}(\cdot\mid \vartheta)$ to be a $\text{Beta}(\alpha,\beta)$ distribution with $\vartheta=(\alpha,\beta)\in \R_{+}^{2}$, and by selecting the marginal prior distributions for the other parameters to be
$\theta\sim \text{Ga}(2,.04)$, 
$\alpha\sim \text{Unif}(0,1)$, 
$\beta\sim \text{Unif}(0,1)$, 
$p\sim \text{Unif}(0,1)$.
These can be thought of as being non informative for $p$, fairly non informative for $\theta$, and strongly informative for $\alpha$ and $\beta$. The prior information on $\alpha$ and $\beta$ is obtained by noting that the average curve shape in the simulated dataset presents steep increments near the extremes of the domain and a central near-plateau. We repeated the experiment with less informative gamma priors for $\alpha$ and $\beta$ (which are also used for the second simulated dataset) and in that case one might need to set initially adaptive ABC thresholds, which start from higher values to increase the chances of obtaining the first accepted ABC sample, and then let them converge to the desired threshold values upon successive acceptances. 
Note that the above hyperpriors assume independence between the  parameters, but this holds only for the first acceptance, since the joint ABC-Metropolis--Hastings acceptance criterion implicitly treats the parameters as correlated. 

The proposal Markov kernel $q(\eta^{*}\mid \eta^{(i)})$ in \eqref{proposal kernel} is set to be given by four truncated normals centered on the last accepted value, with standard deviations 3 for $\theta$ and $.15$ for $p$, $\alpha$, and $\beta$. With symmetric kernels, \eqref{metropolis condition} reduces to the ratio of priors, which further simplifies to 1 when $\pi_{i}$ is uniform. We also set the threshold values to $(\varepsilon_{1},\varepsilon_{2},\varepsilon_{3})=(20,.1, .0003)$, which roughly correspond to $(c_{1},c_{2},c_{3})=(.35,.5,.02)$ in \eqref{c-soglie sim1}.

Finally, $n$ is calibrated to be the integer part of $1/m$, where here $m=\max(\tilde m(F),\text{\emph{tol}})$, $\tilde m(F)$ is the minimum observed jump size in the data and \emph{tol} is a desired tolerance level of jump size approximation to avoid using too many particles. Letting $tol=10^{-3}$, $\tilde m(F)$ results in $2\times 10^{-3}$, so we set $n=500$. Fixing $n$ yields the identifiability of $\theta$, since otherwise different combinations of $(\theta,n)$ could provide the same expected number of  jumps per curve.

We adopt the strategy outlined in Section \ref{sec: abc} and obtain 50000 samples with the MCMC-ABC sampler given in Algorithm \ref{algorithm}, using the first 100 curves in $F$ for computing the summaries $\rho_{j}(F)$ and keeping the last 10 curves as test set for the out-of-sample $h$-step-ahead prediction. We consider three different forecast horizons, for $h=1,3,10$, and evaluate the prediction accuracy with 1000 Monte Carlo forecast samples conditional on $F_{T}$. 
The left column of Figure \ref{fig:sim1-band} shows the $h$-step-ahead forecast estimates for $h=1$ (top), 3 (middle) and 10 (bottom). In each picture the solid line is the true curve to predict, the dashed line is the estimate of the mean posterior predictive, obtained as in \eqref{point estimate}, and the grey region corresponds the 99\% pointwise credible bands. The results provide relatively accurate forecasts in terms of point estimates and thin credible bands for close time horizons, while the increased uncertainty for farther time horizons accounts reasonably for the volatility of the time series. 

\begin{figure}[t!]
\begin{center}
\includegraphics[width=.8\textwidth]{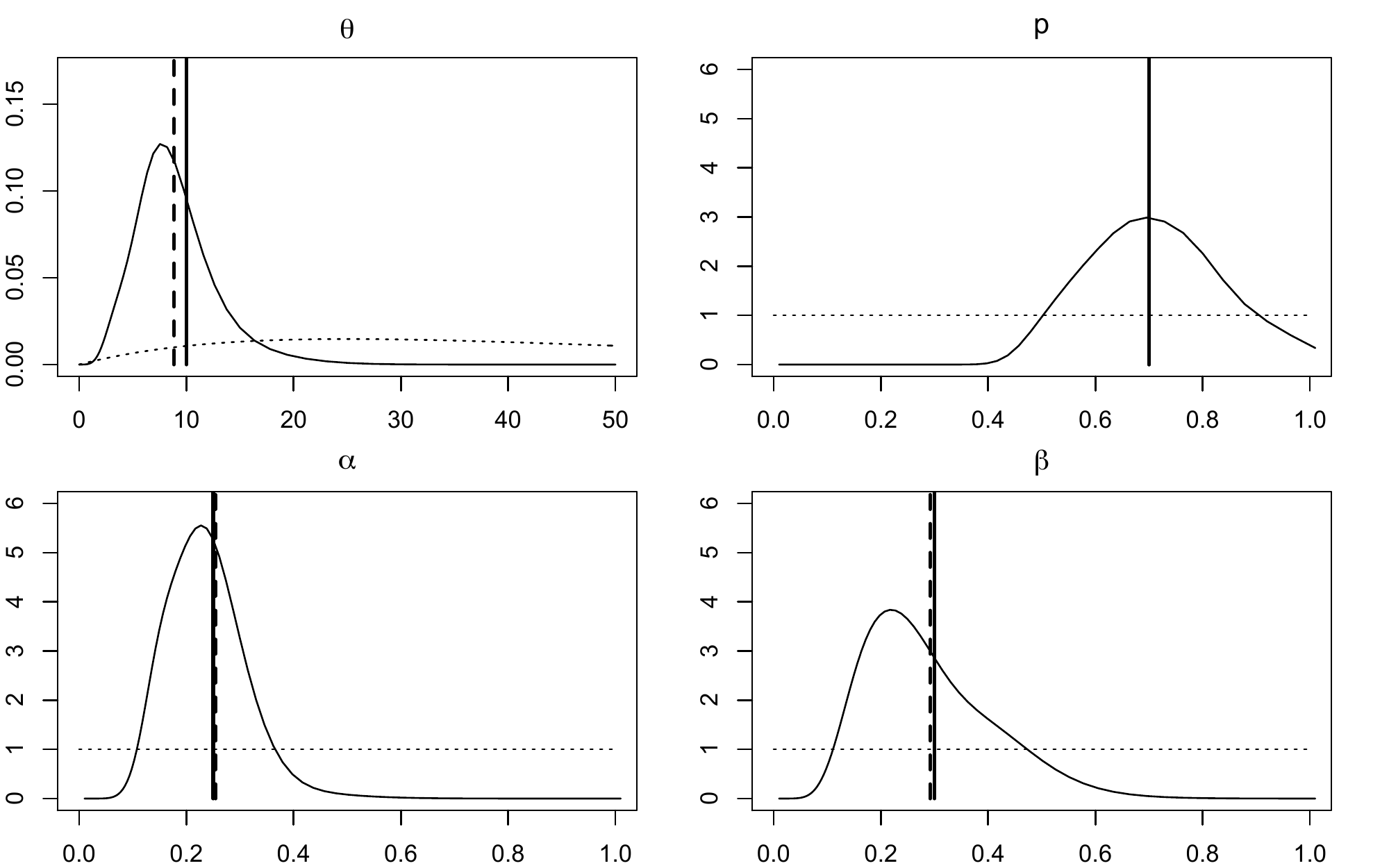}
\end{center}
\begin{quote}
\caption{\scriptsize 
Posterior densities of the model parameters $(\theta,p,\alpha,\beta)$ (solid curves) and posterior means (vertical dashed lines) obtained from the MCMC-ABC samples, against the relative priors (dotted curves) and the true parameter values (vertical solid lines).
}\label{fig:sim1-par}
\end{quote}
\end{figure}
Figure \ref{fig:sim1-par} shows the posterior densities of the model parameters $(\theta,p,\alpha,\beta)$ (solid curves) and the posterior means (vertical dashed lines) along with the priors (dotted curves) and the true parameter values (vertical solid lines). The posterior densities of the parameters are obtained with a standard kernel smoothing technique applied to the MCMC-ABC samples.

We then consider a model misspecification, and assume the data are generated by a functional autoregression of the form
\begin{equation*}
F_{t}=aF_{t-1}+(1-a)F_{\varepsilon},\quad \quad a \in (0,1),
\end{equation*} 
with $F_{0}=F_{\varepsilon}$. The curve at time $t$ is a convex linear combination of the previous curve and a noise term $F_{\varepsilon}$ defined to be the empirical cdf of 20 samples from a $\text{Beta}(\alpha,\beta)=\text{Beta}(5,3)$. 
For the inference we set
$\theta\sim \text{Ga}(2,.04)$, 
$\alpha\sim\text{Ga}(2,.25)$,
$\beta\sim \text{Ga}(2,.25)$,
$p\sim \text{Unif}(0,1)$. The proposal Markov kernel in the ABC sampler is the same as for the previous illustration, this time with standard deviations 3 for $\theta$, $\alpha$ and $\beta$ and $.1$ for $p$. We set the threshold values to $(\varepsilon_{1},\varepsilon_{2},\varepsilon_{3})=(25,.095, .008)$, which roughly correspond to $(c_{1},c_{2},c_{3})=(.15,1,.6)$ in \eqref{c-soglie sim1}. The forecast results are presented in the right column of Figure \ref{fig:sim1-band}.
The model performs well by producing reliable out-of-sample predictions, demonstrating robustness to a model misspecification. 

With the aim of having a quantitative measure of the performance alongside the visual results, Table \ref{tab:l2curve-sim} reports the $L_2$ distances between predicted  and  true curves for both simulation studies, confirming that the quality of the prediction slightly deteriorates for increasing $h$ but remains acceptable. 
The better performance in the misspecified case can be related to the lower volatility of the time series with respect to the first simulated dataset.

\begin{table}
   \centering
   \begin{tabular}{|c|c|c|}\hline
$h$    & Sim 1 & Sim 2 \\\hline
    1 &0.0101 & 0.00093\\
3 & 0.0137& 0.00153\\
10 & 0.0453 & 0.00974 \\\hline
   \end{tabular}
   \vspace{2mm}
\caption{$L_2$ distances between $h$-step-ahead predictions and true curves (dashed and solid lines in Fig.~\ref{fig:sim1-band}).}
   \label{tab:l2curve-sim}
\end{table}


\section{Application to natural gas market}\label{sec: appl}

\subsection{The natural gas virtual balancing platform}
\label{sec:context}

In the last decade, many European natural gas markets have undergone radical changes, such as the legal splitting of pipeline managers and gas shippers, and the introduction of legislation for obligatory third party access to transmission, distribution, and storage of natural gas capacity \citep{Eurosparliament}. 
Aimed at favouring a liberal market, such measures also brought several new logistic challenges. In Italy, for example, the control of the national pipeline network is managed by Snam s.p.a., an independent actor from the several gas traders that inject natural gas into the common pipeline network. Its role is to compensate injections and consumptions via storage or other measures. In fact, the risk of possible unbalance of the network is assigned to each shipper which has to daily predict and communicate to Snam its injection and consumption forecasts, on which a penalty is payed for any positive or negative unbalance. With the idea of having a self balancing system, in December 2011 the policy maker introduced the natural gas balancing platform, a virtual market where gas operators and traders buy and sell natural gas in order to balance the common pipeline.  

\begin{figure}
\begin{center}
\subfigure[]{\includegraphics[width=.6\textwidth]{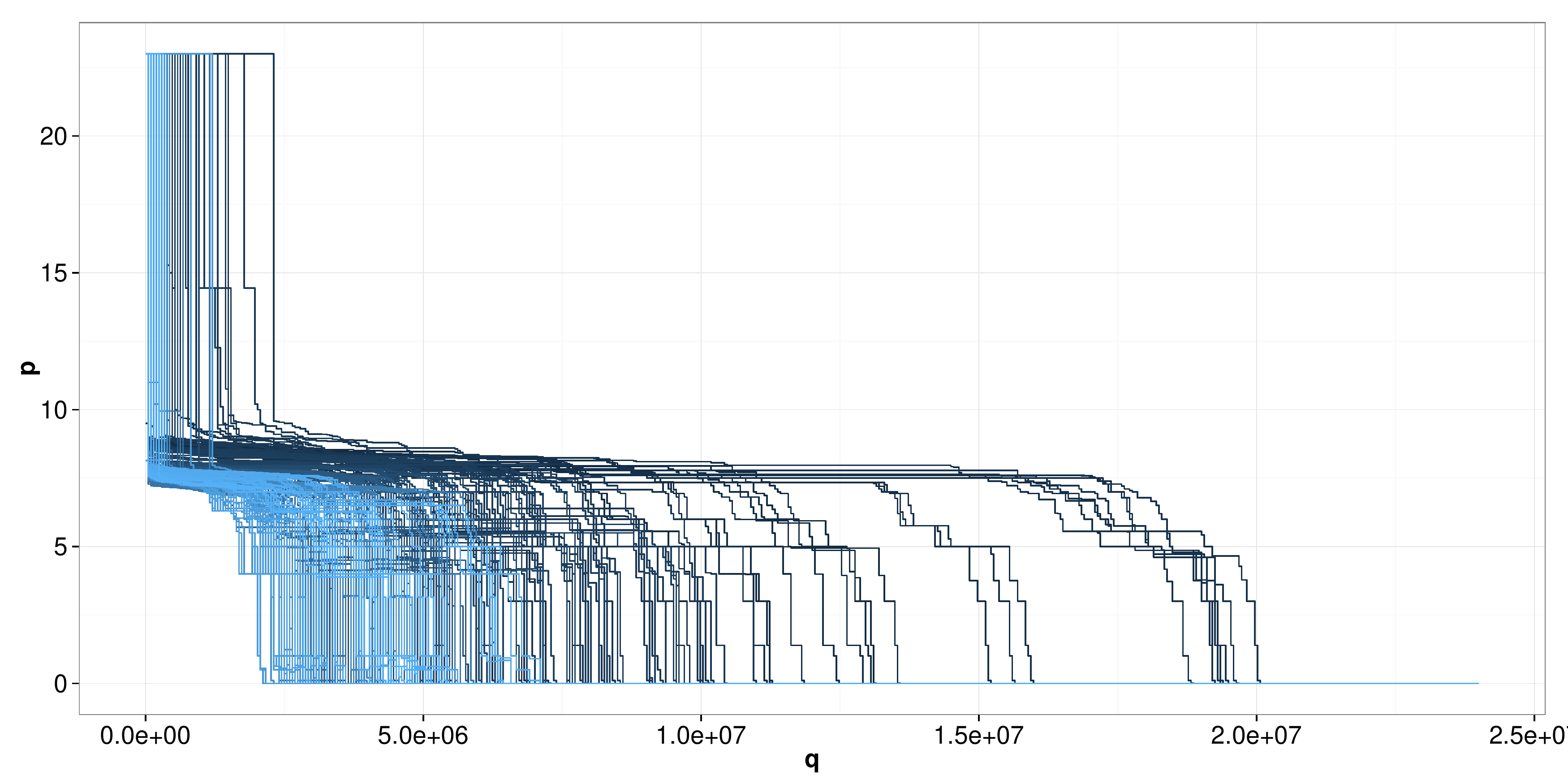}}
\subfigure[]{\includegraphics[width=.6\textwidth]{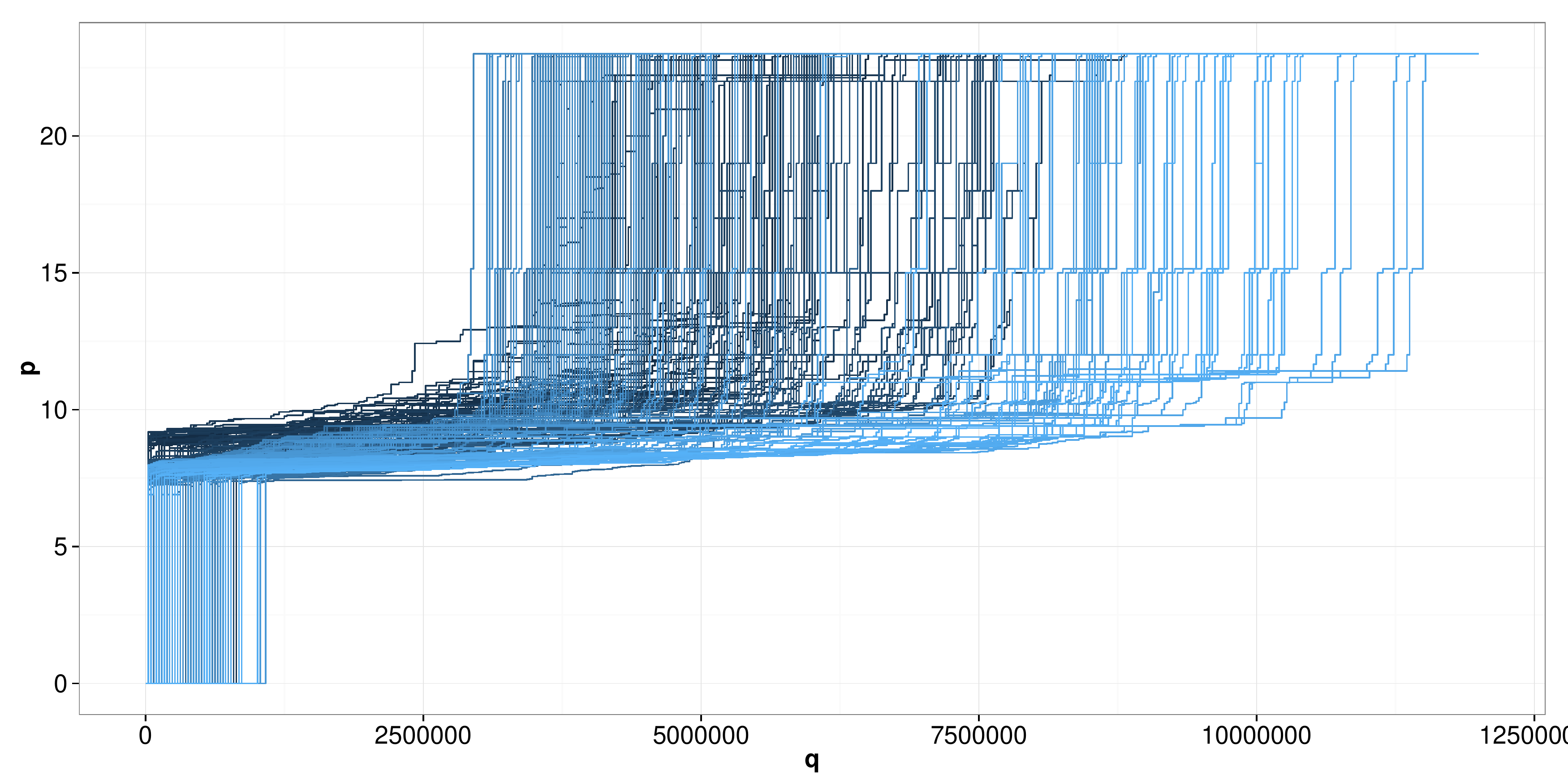}}
\subfigure[]{\includegraphics[width=.6\textwidth]{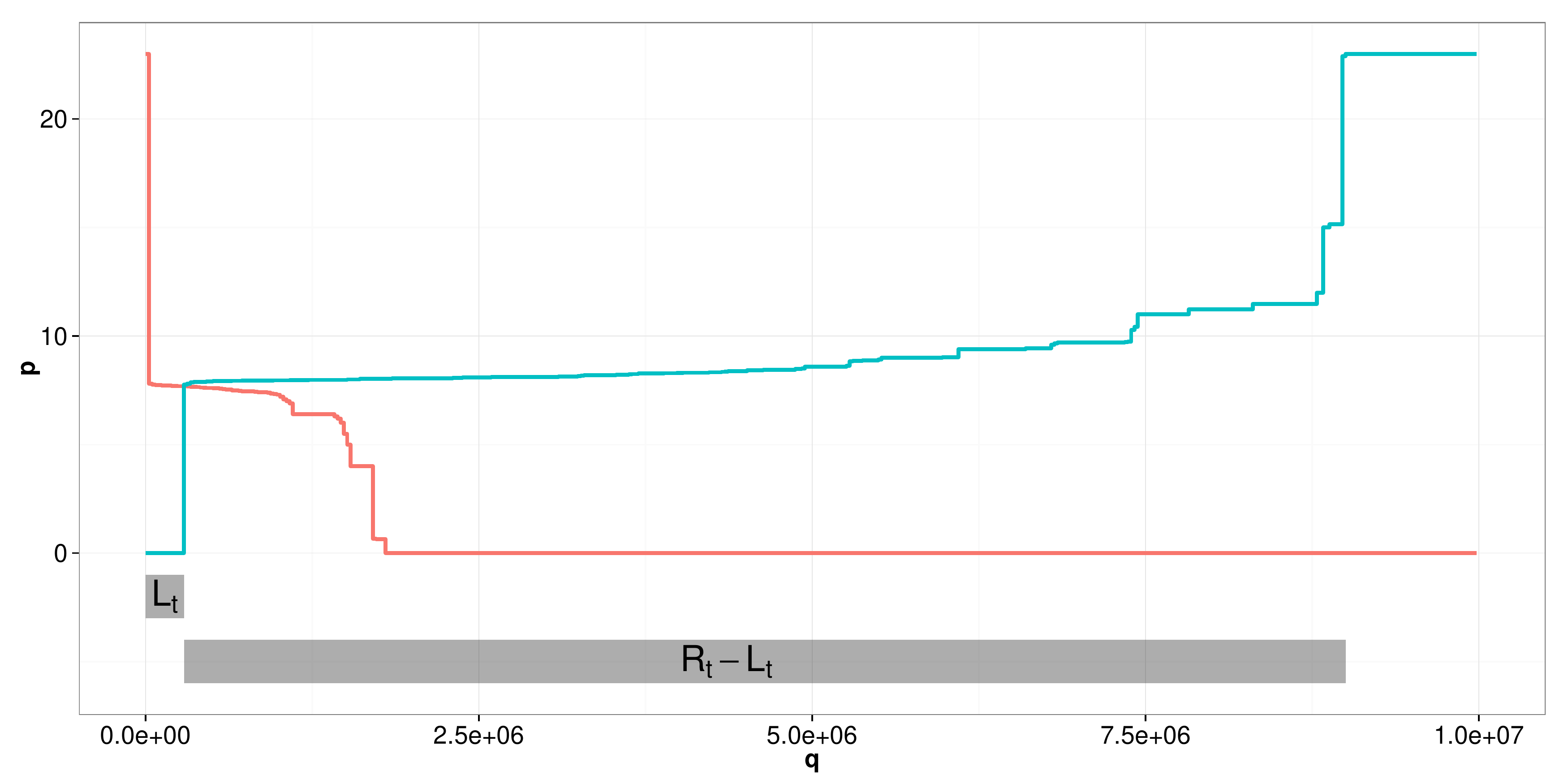}}
\end{center}
\caption{\scriptsize 
Functional time series from the Italian natural gas market dataset, given by demand \emph{(a)} and offer \emph{(b)} curves, brigther colours indicating more recent curves. Panel \emph{(c)} highlights the leftmost ($L_t$) and rightmost ($R_t$) jump locations, used in \eqref{eq:phi}, for an offer curve.
}\label{fig: real data}
\end{figure}

The virtual balancing platform works as follows \citep[the interested reader is referred to][for more details]{GME}. Every day Snam submits a demand bid or supply offer for a volume of gas corresponding to the overall imbalance of the system, while the operators submit demand bids and supply offers for the storage resources they have available. 
Demand bids are sorted by price in decreasing order, and viceversa for supply offers, and the demand and supply curves are obtained as the cumulative sums of the respective quantities in gigajoules (GJ). The resulting offer (resp.~demand) curves are monotone increasing (resp.~decreasing) step functions with positive domains and bounded image. See Figure \ref{fig: real data}.  With an auction mechanism, offers on the left of the intersection of the demand and supply curves are accepted and the transactions are carried out at the intersecting price. Hence, bidding a demand (resp.~offer) at the maximum (resp.~minimum) permitted price (enforced to be 0 and 23 Euros/GJ), allows Snam to always be at the left of the intersection, which ultimately determines the overall system balancing by Snam selling (resp.~purchasing) the gas excess (resp.~deficit) to other shippers at the exchange price.  

Here it is worth emphasising the difference between the single price and quantity values of a general bid and the particles system introduced in Section \ref{sec: model}. Bids are, indeed, not samples from the resulting increasing step functions since the latter cannot be considered as the data generating mechanism, despite they share the same characteristics of a discrete cdf. Hence the bids cannot be directly identified with particles. 

While balancing the common pipeline network, this regime change has created new opportunities for traders. 
From a speculative viewpoint, one could take advantage of the platform mechanism for buying natural gas at a lower price or selling exceeding gas at a higher price, relative to their benchmark supplying indices. This requires suitable bidding strategies based on reliable predictions of the quantities at stake. In this respect, parametric time series forecasting can be severely limitative, as the trader needs to predict the effect of his own bid on the future intersection of demand and supply curves. Making available an estimate of the entire tomorrow's curve helps predicting both the exchange price and the total exchanged quantity of gas, and is therefore of crucial importance for predicting the price dynamics.

\begin{figure}[t]
\begin{center}
\subfigure[]{\includegraphics[width=.42\textwidth]{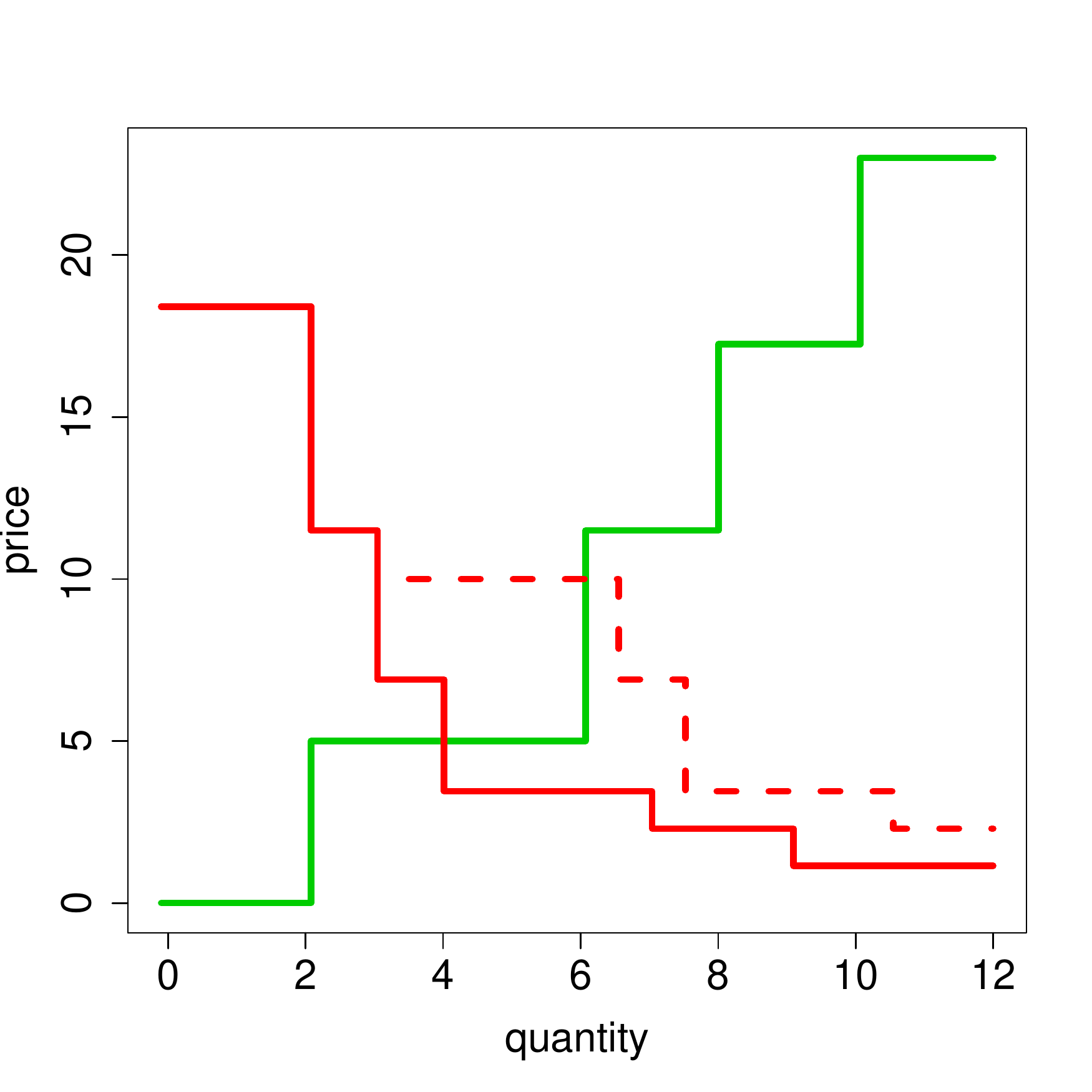}}
\subfigure[]{\includegraphics[width=.42\textwidth]{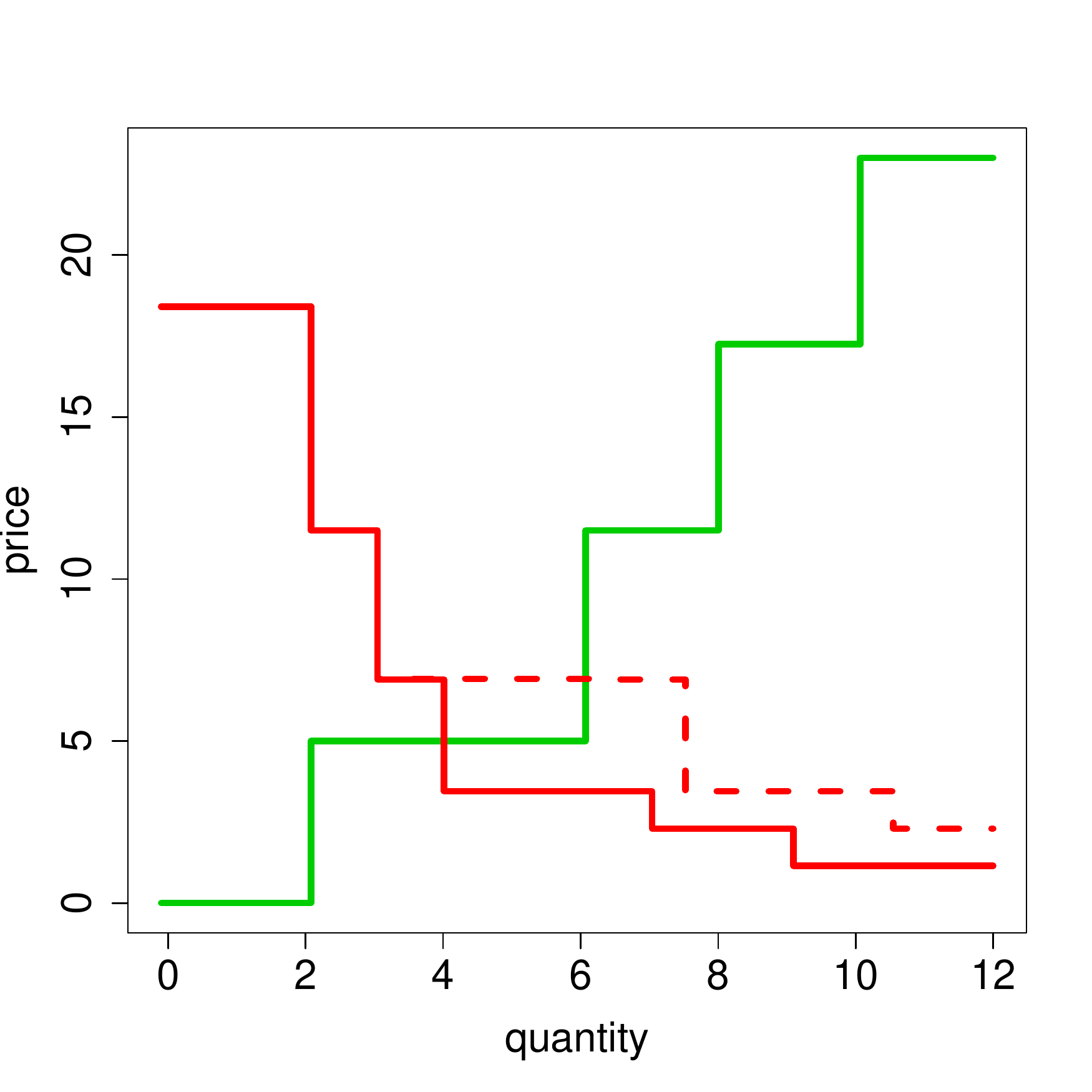}}
\end{center}
\begin{quote}
   \caption{\scriptsize Stylised trading example. The intersection of the estimated  curves (solid lines) predicts the exchange price. Different bidding strategies modify the curves (dashed curves) and allow to forecast the effect on the resulting exchange price.}\label{fig:toy}
\end{quote}
\end{figure}

To emphasise this important aspect, consider the following toy example from the perspective of a trader. 
Suppose it is convenient for trader to buy at a price not exceeding 8 Euros/GJ, and he is provided the curves forecast, given by Figure~\ref{fig:toy} (solid curves in both panels). The intersection predicts the price to be 5 Euros/GJ, hence any offer with price between 5 and 8 will be convenient for the trader and likely to be accepted, depending on the resulting actual exchange prices, whereas offers below the actual exchange price will be refused. Here different actions by the trader will affect differently the actual curve and thus the price. For example, if the trader offers to buy 3.5 GJ at 10 Euros, its offer will be highly likely accepted but the exchange price will also likely rise. For example Figure~\ref{fig:toy}-(a) shows that if the trader offers to buy 3.5 GJ at 10 Euro, the exchange price will be 10 Euros, no longer convenient for the trader. If instead the trader offers to buy 3.5 GJ at 7 Euros, the offer will also be likely accepted but the resulting exchange price will likely fall between 5 and 7 and be more convenient for the trader, as in Figure~\ref{fig:toy}-(b). This stylised example shows that \emph{what-if}-type simulations on the predicted curve can lead to conclusions which are unavailable if based on the univariate price predictions only.

\subsection{Dataset and model specification}\label{subsec: dataset}

We focus on the first available data relative to the Italian Natural Gas Balancing Platform, which run from January 2012 to December 2012, of which the last month is removed and used as a test set for out-of-sample prediction.
The original data consist of a daily table where each row represents an awarded bid. From this dataset we build the offer (resp.~demand) curves by sorting the selling bids by price in increasing (resp.~decreasing) order and obtaining the value of the quantities by cumulating each single awarded quantity. See Fig.~\ref{fig: real data} (a)-(b). This information is made available with a one-week delay, so in this specific framework the 8-step-ahead forecast is the meaningful estimate. We also consider 1- and 3-step-ahead forecasts for illustration. 

The curves can be normalised to take values in $[0,1]$ by exploiting the maximum admitted price of 23 Euros and the demand curves can be reversed to work with non decreasing functions. We thus obtain two functional time series of monotone increasing step functions with positive domain and taking values in $[0,1]$.

Despite the one week lag for the full data availability, every day the pipeline network manager Snam declares the sign and the magnitude of the next day's imbalance. As previously discussed, Snam daily submits a unique bid which determines the location of the first step of one of the two curves. If the total network imbalance  is positive, Snam submits a single demand bid with maximum price and quantity equal to the imbalance, which makes the location of the first steps of the demand and offer curves equal to the overall imbalance and to zero, respectively. If the total imbalance is negative, Snam submits a single supply offer with minimum price and quantity equal to the imbalance, which makes the location of the first steps of the demand and offer curves equal to zero and the overall imbalance, respectively. Thus, in practice, the location of the first jump of both curves is known one day in advance, and therefore the respective time series are not considered here as part of the inferential goals. 

Concerning the location of the last jump, if some market player is willing to avoid entering the auction, a way to do this consists in bidding a demand (resp.~offer) at the minimum (resp.~maximum) permitted price. This in turn creates the last jump of the curves, beyond which one can find other bids which do not influence the exchange price nor the relevant region of the curve shape. 
This somewhat exogenous determination of the first and last curve jumps justifies modelling them independently of the overall curve dynamics. To this end we modify \eqref{curve t-indexed} to have
\begin{equation}
F_{t}(x)=
\frac{1}{n}\sum_{i=1}^{n}\ind(Y_{i}(t)\le x), \quad Y_i(t) = L_t + (R_t-L_t)X_i(t),
\label{eq:phi}
\end{equation}
where $\{L_t\}_{t=0,\ldots,T}$ and $\{R_t\}_{t=0,\ldots,T}$ are one dimensional time series representing the locations of the leftmost and rightmost jumps in each curve respectively, with $L_{t}<R_{t}$ for all $t$. See Figure \ref{fig: real data}-(c). 
We will model the latter and consider the former as known as discussed above.
Denoting by $F_t^D(x)$ and $F_t^S(x)$ the demand and supply curves, normalise them to yield
\begin{equation}\label{curve nomalizzate}
F_{t}^d(x) = F_t^D\left( \frac{x-L_t^d}{R_t^d - L_t^d}\right),\quad 
F_{t}^s(x) = F_t^S\left( \frac{x-L_t^s}{R_t^s - L_t^s}\right).
\end{equation}
We model $F_{t}^d$ and $F_{t}^s$ as in \eqref{curve nomalizzate} and then transform back according to the known $L_t$ and estimated $R_{t}$. 

Both series $R_{t}^d$ and $R_{t}^s$ present traits of non stationarity. We consider then the differentiated log series, which exhibits stationary behaviour and allows for strictly positive forecasts. To this aim, we let
\begin{equation*}
r_t = \rho r_{t-1} + \epsilon, 
\quad \epsilon \sim N(0, \sigma^2),
\quad r_t = \log(R_{t+1}) - \log(R_{t}),
\end{equation*} 
and assign non informative priors $\rho\sim N(0, 1000)$ and $\sigma^2\sim \mbox{I-Ga}(0.01, 0.01)$, where I-Ga denotes an inverse gamma distribution. The posterior computation is performed via MCMC sampling using JAGS \citep{JAGS}. 

To estimate the curve trajectories, we implement our MCMC-ABC algorithm of Section~\ref{sec: abc} to collect 10000 ABC samples, on the base of which 1000 forecast samples are generated conditional the last curve. We set $n=500$, 
$\theta\sim \text{Ga}(2,.04)$, 
$\alpha\sim\text{Ga}(2,20)$,
$\beta\sim \text{Ga}(2,20)$ and
$p\sim \text{Unif}(0,1)$. The proposal Markov kernel in the ABC sampler is the same as for the simulation study with standard deviations 3 for $\theta$ and .05 for $p$, $\alpha$ and $\beta$, and the threshold values are set equal to $(\varepsilon_{1},\varepsilon_{2},\varepsilon_{3})=(25,.07, .01)$, which roughly correspond to $(c_{1},c_{2},c_{3})=(.78,.2,.4)$ in \eqref{c-soglie sim1}.

\subsection{Results}

\begin{figure}
   \centering
\subfigure[$h=1$]{\includegraphics[page=1, width=.35\textwidth]{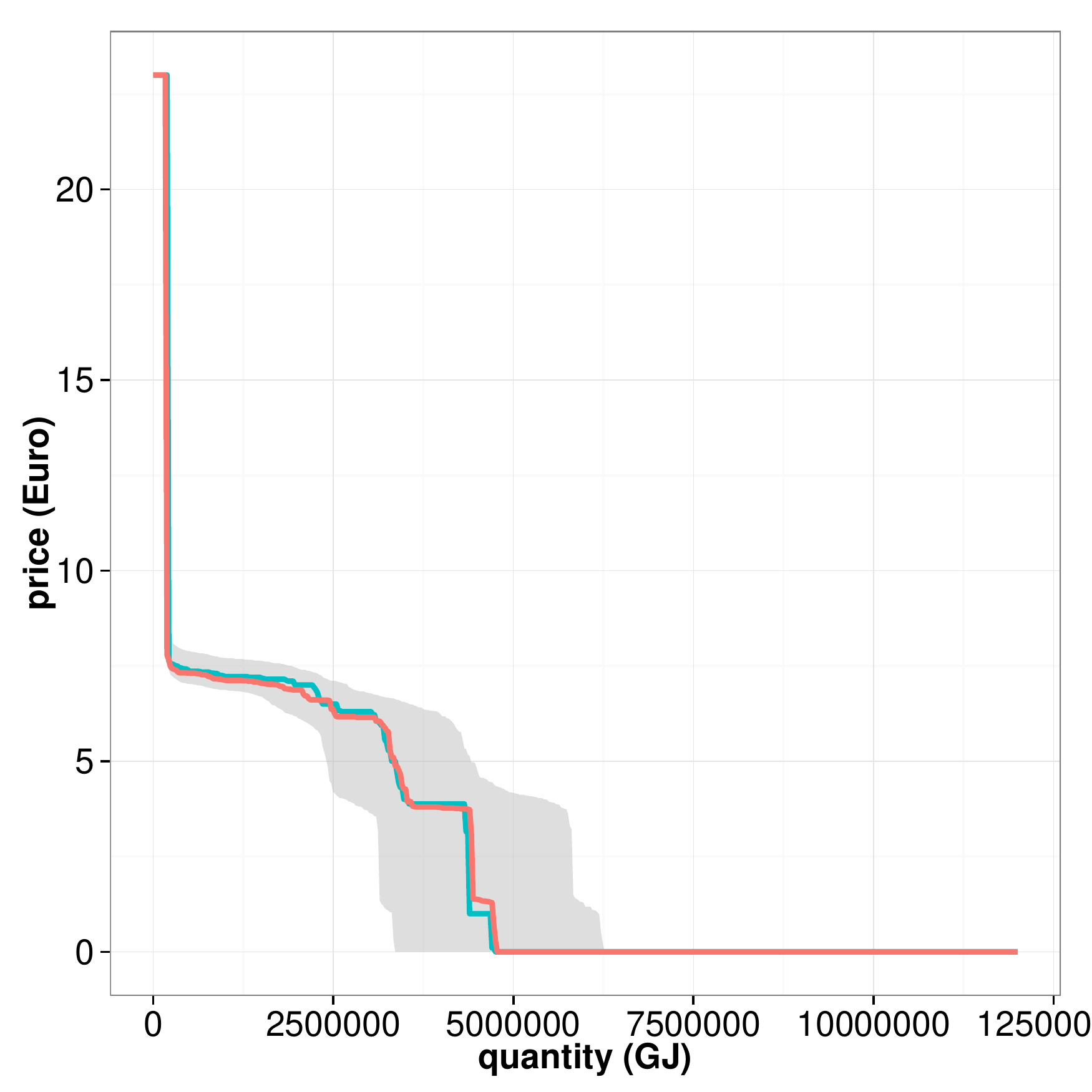}}
\subfigure[$h=1$]{\includegraphics[page=2, width=.35\textwidth]{real_forecast_h1}}
\subfigure[$h=3$]{\includegraphics[page=1, width=.35\textwidth]{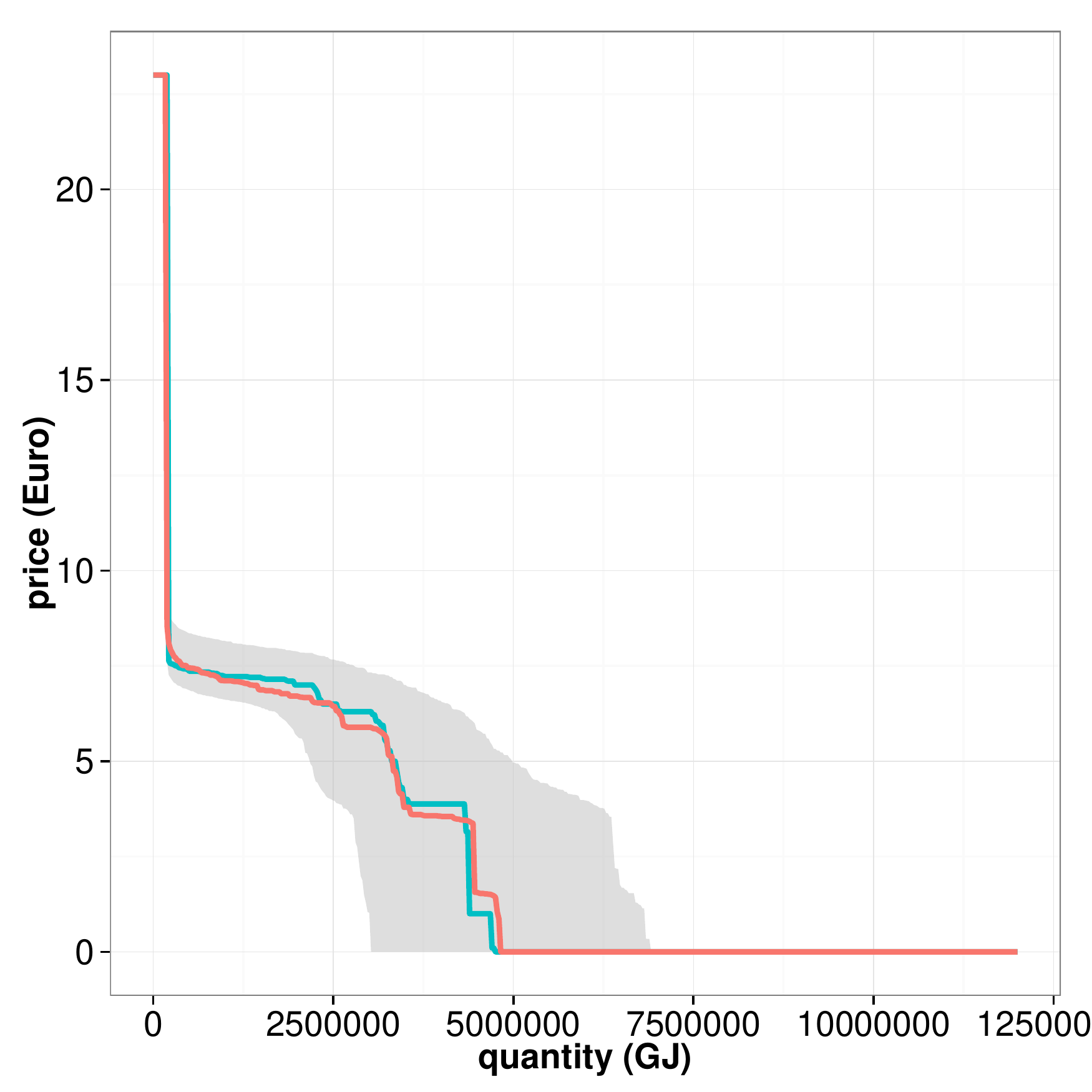}}
\subfigure[$h=3$]{\includegraphics[page=2, width=.35\textwidth]{real_forecast_h3}}
\subfigure[$h=8$]{\includegraphics[page=1, width=.35\textwidth]{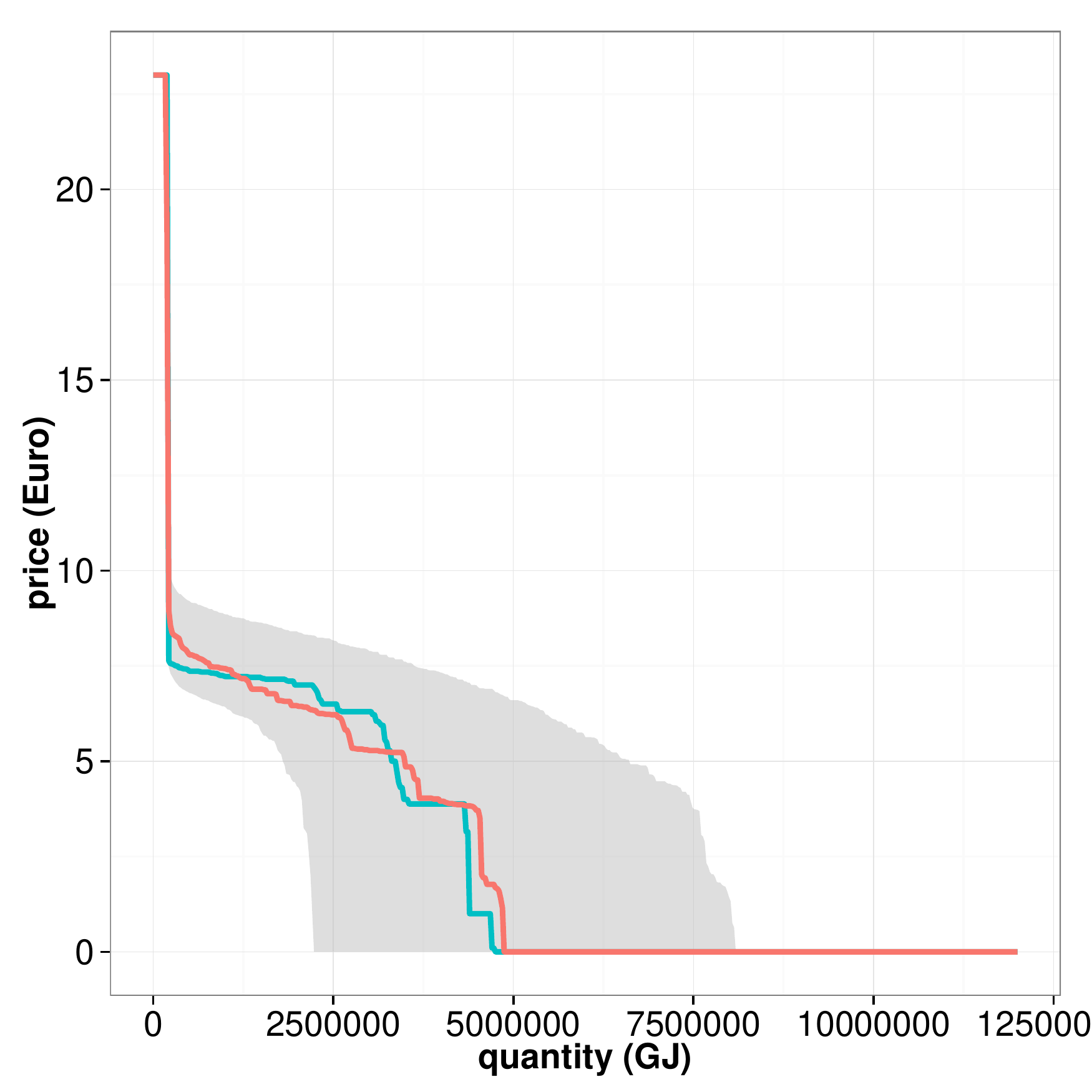}}
\subfigure[$h=8$]{\includegraphics[page=2, width=.35\textwidth]{real_forecast_h8}}
   \caption{\scriptsize  True demand (left column) and offer (right column) curves for the 31st of December 2012 with predictive point estimates and 95\% credible bands using a $h$-step-ahead posterior predictive distribution with $h=1$ (top row), $h=3$ (middle row), and  $h=8$ (bottom row).}
   \label{fig:31decembre}
\end{figure}

Figure~\ref{fig:31decembre} shows an instance of out-of-sample predictions for December 31st, 2012. Each panel shows the true curve (cyan) and its point estimate (red), based on \eqref{point estimate}, along with pointwise 95\% credible bands (grey region). 
The results are consistent with the simulations in giving reliable point $h$-step-ahead forecasts and credible bands, the latter manifesting  increasing uncertainty as $h$ increases. The estimation is slightly less precise for the offer curves. In both cases, offer and demand, the increasing width of the credible bands for higher values of the domain is due to the variability of the last jump location series $R_t^s$ and $R_t^d$, which clearly impact the curves scale on their codomain. A more sophisticated model for $R_t^s$ and $R_t^d$ could give a better performance in terms of width of the final credible bands but, as mentioned in Section \ref{subsec: dataset}, this does not have a relevant impact on the application at hand. The output of the estimation can be used to forecast the impact of trading strategies on the curves intersection, which typically occurs in the first part of their domain, as discussed in the stylised example of Section~\ref{sec:context}. 

Table~\ref{tab:l2curve} summaries the estimation performance throughout the test set by reporting the mean $L_2$ distances between the predicted curves and true one computed along the test set, here December 2012. The measures are calculated both on the original scale, i.e.~$F:[0,1.2 \times 10^7]\rightarrow [0,23]$ as in Figure \ref{fig: real data}, and on a normalized scale, \mbox{i.e.~$F:[0,1]\rightarrow [0,1]$}, for which the $L_{2}$ distance also ranges in $[0,1]$.  As expected, the quality of prediction slightly deteriorates for increasing $h$ but remains overall satisfactory.

\begin{table}[t]
   \centering
   \begin{tabular}{|c|cc|cc|}\hline
    & \multicolumn{2}{c|}{Original scale} & \multicolumn{2}{c|}{Normalized scale} \\
   $h$ & Offer & Demand & Offer & Demand \\\hline
    1 & 1,347,221 & 1,351,971 & 0.0049 & 0.0049\\
3 & 1,499,623 & 1,445,809 & 0.0054 & 0.0052\\
8 & 2,141,462 & 1,944,311 & 0.0078 & 0.0070\\\hline
   \end{tabular}
   \vspace{2mm}   
   \caption{\scriptsize Mean $L_2$ distances between pointwise $h$-step-ahead predictions and real curves computed for the test set (December 2012).}
   \label{tab:l2curve}
\end{table}

\begin{figure}[t]
   \centering
Price forecast from functional prediction\\
\subfigure[$h=3$]{\includegraphics[page=1, width=.3\textwidth]{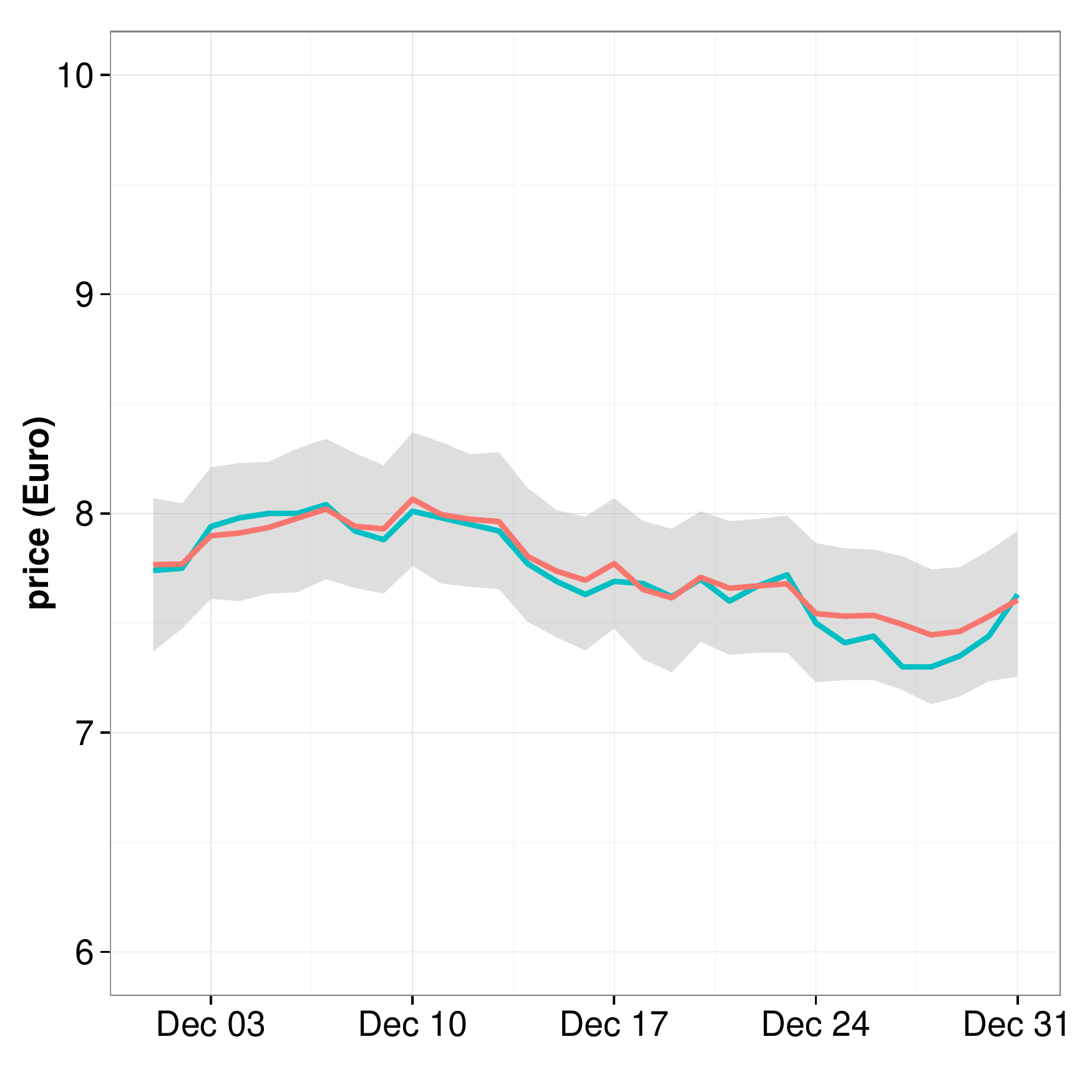}}
\subfigure[$h=8$]{\includegraphics[page=2, width=.3\textwidth]{prices}}
\subfigure[$h=1$]{\includegraphics[page=3, width=.3\textwidth]{prices}}\\
Price forecast from univariate autoregressive model\\\subfigure[h=1]{\includegraphics[page=1, width=.3\textwidth]{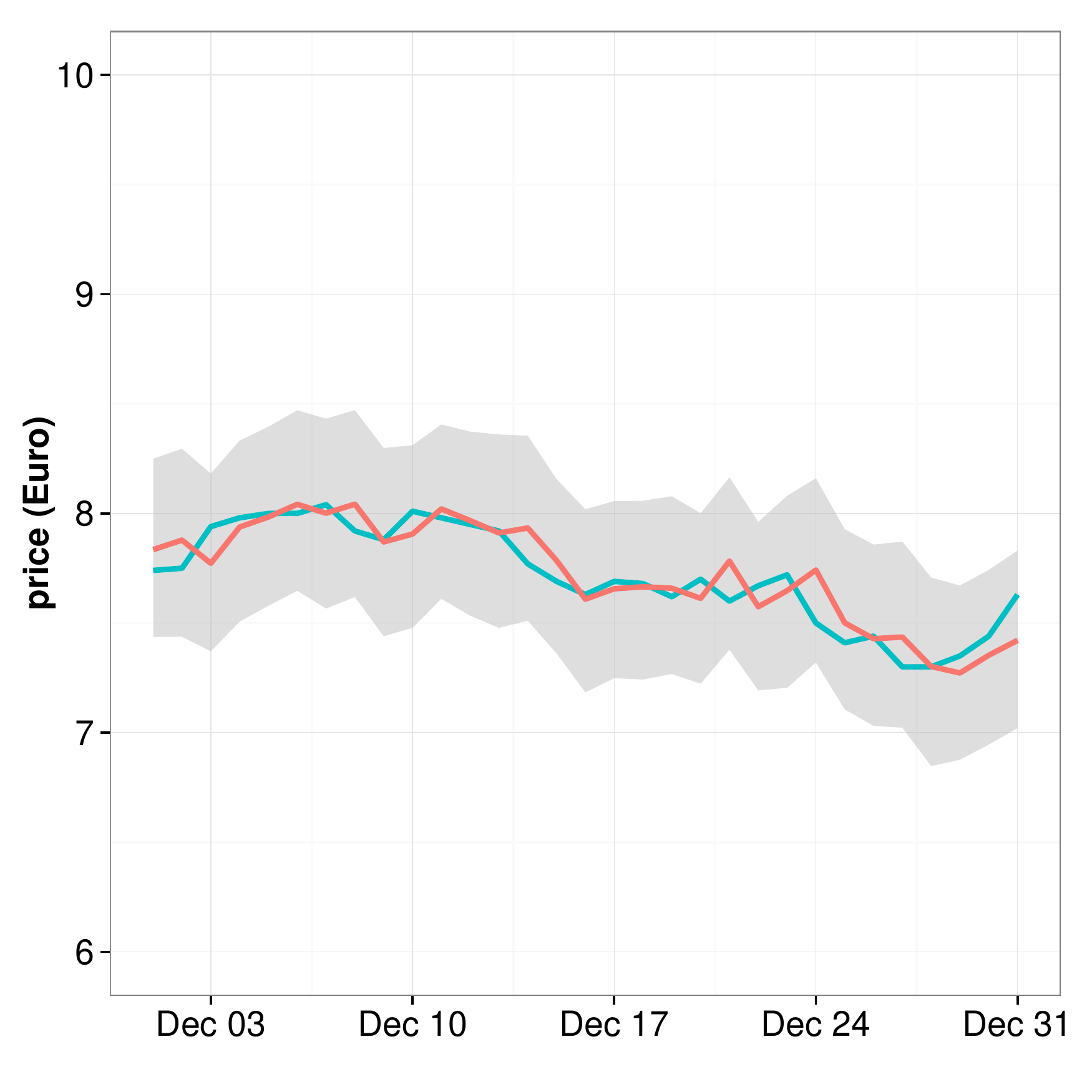}}
\subfigure[$h=3$]{\includegraphics[page=2, width=.3\textwidth]{prices_arprice}}
\subfigure[$h=8$]{\includegraphics[page=3, width=.3\textwidth]{prices_arprice}}
\caption{\scriptsize  Comparison of true price (cyan) and its $h$-step-ahead forecast (red) along with pointwise 95\% credible bands (grey), for $h=1$ (left column), $h=3$ (middle column), and $h=8$ (right column). Top row: price obtained as intersection of the functional predictions of demand and supply curves. Bottom row: price obtained by fitting a univariate autoregressive model.}
   \label{fig:priceforecast}
\end{figure}

As a byproduct, forecasts of associated features of interest can be derived from the functional predictions, in this context the equilibrium exchange price. Inference on these quantities is usually addressed by means of univariate parametric temporal models. Here, a Bayesian nonparametric price forecast is obtained as the intersection point of the posterior predictions of offer and demand curves. Figure~\ref{fig:priceforecast} (top row) shows the predictive mean (red) and 95\% credible pointwise bands for the exchange price (grey), along with true values (cyan) for December 2012.

We compare these results with those obtained fitting a simple auto regressive model to the differentiated series of prices. Specifically, letting $p_t$ be the price at day $t$ and $\tilde{p}_t = p_{t+1} - p_{t}$, we fit 
\begin{equation*}
\tilde{p}_t = \psi \tilde{p}_{t-1} + \epsilon_t, 
\quad \epsilon_t \sim N(0, \sigma^2),
\end{equation*} 
assuming $\psi\sim N(0, 1000)$ and $\sigma^2\sim \mbox{I-Ga}(0.01, 0.01)$ via MCMC sampling using again JAGS. As before we consider the data of the first 11 months of 2012 as training test and the last month as test set. The bottom row of Figure~\ref{fig:priceforecast} shows the resulting Monte Carlo predictive mean and associated 95\% credible pointwise bands. 
Table~\ref{tab:prezzicfr} reports the root mean squared error for the $h$-step-ahead predictions ($h=1,3,8$) for both approaches.  
Although not specifically tailored for this univariate dynamic inference, the proposed model exhibits a good performance in the price prediction compared. Incidentally, this provides an interesting example of a free byproduct of a nonparametric method, developed for a more general purpose, specified to a particular goal usually addressed by ad hoc procedures.

\begin{table}
   \centering
   \begin{tabular}{|c|c|c|}\hline
    & Intersection of & Univariate \\
   $h$ &  functional time series &  time series  \\\hline
    1 & 0.07 & 0.10 \\
	3 & 0.09 & 0.15 \\
	8 & 0.19 & 0.23 \\\hline
   \end{tabular}
   \vspace{2mm}   
      \caption{\scriptsize Root mean squared error for the $h$-step-ahead exchange price forecast over the test set (December 2012), obtained as the intersection of the functional predictions and by fitting an autoregressive model.}
   \label{tab:prezzicfr}
\end{table}

%
\section{Concluding remarks}\label{sec: extension}

Anticipating price dynamics can be extremely important in a variety of economic contexts, for optimising resources and operating strategies. In energy markets, traders have an advantage if they can forecast the impact of their own bidding strategies on the estimated price, which is in turn determined by the aggregate predicted future demand and supply curves. Making this estimates available allows market operators to judge such impact well beyond similar evaluations based on predictions which rely on univariate parametric models. 
Here we have proposed a simple Bayesian nonparametric model that shows a satisfactory performance both at the functional and at the implied univariate level, thus constituting a reliable instrument in the hands of the operators. 

Modifications of the model can be useful for similar estimation problems in different frameworks. The transformation of the particles in \eqref{eq:phi} can be clearly extended to account for different curves characteristics, still with the aim of modelling separately the temporal evolution of the curves' shape, directly handled by the particle system, and their location and scale. Another extension consists in relaxing the Markovianity requirement in the construction of the particle system. For instance, the particles can be resampled from vectors older than the previous one, with decreasing probability of sampling from older curves. This would provide an autoregressive version of the model, whose implementation and performance will be explored elsewhere.


%


\end{document}